\documentclass[aps,prd,eqsecnum,preprint,tightenlines,nofootinbib]{revtex4-1}
\usepackage{graphicx,latexsym}

\newcommand{\bea}{\begin{eqnarray}}
\newcommand{\eea}{\end{eqnarray}}
\newcommand{\be}{\begin{equation}}
\newcommand{\ee}{\end{equation}}
\newcommand{\ub}[1]{\underline{#1}}

\begin{document}

\title{The Casimir effect in light-front
quantization%
\footnote{Based on a talk contributed to the
Lightcone 2014 workshop, Raleigh, North Carolina, 
May 26-30, 2014.}
}

\author{J.R. Hiller}
\affiliation{Department of Physics \\
University of Minnesota-Duluth \\
Duluth, Minnesota 55812}

\date{\today}

\begin{abstract}
We show that the standard result for the Casimir force between 
conducting plates at rest in an inertial frame can be computed 
in light-front quantization.  This is not the same as light-front 
analyses where the plates are at ``rest'' in an infinite momentum 
frame.  In that case, Lenz and Steinbacher have shown that the 
result does not agree with the standard result for plates at rest.  
The two important ingredients in the present analysis are a 
careful treatment of the boundary conditions, inspired by the 
work of Almeida {\em et al}.\ on oblique light-front coordinates, and 
computation of the ordinary energy density,  rather than the 
light-front energy density.
\end{abstract}

\maketitle

%%%%%%%%%%%%%%%%%%%%%%%%%%%%%%%%%%%%%%%%%%%%%%%%%%%%%
\section{Introduction} \label{sec:intro}
%%%%%%%%%%%%%%%%%%%%%%%%%%%%%%%%%%%%%%%%%%%%%%%%%%%%%

The Casimir effect is a force between conducting plates that arises
from the exclusion of vacuum modes by boundary conditions at the plates~\cite{Casimir}.
The vacuum energy density between the plates differs from the free density.
This defines a separation-dependent effective potential for the plates,
and variation with respect to the separaton yields the force.
The necessary energy density is obtained from a sum over the allowed modes.
The analysis is simplified by working with a massless scalar field 
and  periodic boundary conditions, in place of the complexity of 
quantum electrodynamics.

Contrary to what has been thought, the standard result for 
the Casimir force between conducting plates at rest in an 
inertial frame {\em can} be computed in light-front quantization~\cite{LFCasimir}.  
This is not the same as light-front analyses where the plates are 
at ``rest'' in an infinite momentum frame~\cite{Lenz}, where the
result does not agree.  Placement of the plates in the correct
frame is critical.

The light-front analysis has two important ingredients.  One
is a careful treatment of the boundary conditions,
inspired by the work of Almeida {\em et al}.~\cite{Almeida} on oblique 
light-front coordinates.  The other is the computation of the 
ordinary energy density, rather than the light-front energy density.
The key point here is to focus on the physics; calculations of the
same physical effect in different coordinate systems must yield
the same result.

We define light-front coordinates~\cite{Dirac,DLCQreviews} as 
$x^+=t+z$ for time and $\ub{x}=(x^-,\vec{x}_\perp)$ for space,
with $x^-\equiv t-z$ and $\vec{x}_\perp=(x,y)$.  The light-front
energy is $p^-=E-p_z$, and the light-front momentum is $\ub{p}=(p^+,\vec{p}_\perp)$,
with $p^+\equiv E+p_z$ and $\vec{p}_\perp=(p_x,p_y)$.  The 
mass-shell condition $p^2=m^2$ becomes $p^-=(m^2+p_\perp^2)/p^+$.

The natural choice for the analysis of plates at rest in an inertial
frame would be equal-time coordinates, with periodicity of the field
along one spatial direction.  The ``natural'' situation for
light-front coordinates is to have spatial periodicity in $x^-$,
but this corresponds to plates moving with the speed of light.
Lenz and Steinbacher considered this arrangement~\cite{Lenz}.
Almeida {\em et al}.~\cite{Almeida} used oblique light-front coordinates
with $\bar{x}^0=t+z$ and $\bar{x}^3=z$, and
boundary conditions periodic in $z$, to obtain the correct result.
Both attempts implied that light-front quantization is deficient
in some way; instead, light-front coordinates are just harder to use.

In addition to using boundary conditions appropriate to
plates at rest in an inertial frame, one must calculate 
the true vacuum energy.  This is not the light-front energy $p^-$.
The equal-time energy $E$ is what determines the effective potential 
and, therefore, the force.  Other examples of where this choice
matters can be found in the 
variational analysis of $\phi^4$ theory~\cite{Vary} and
the calculation of thermodynamic partition functions.  In particular,
a partition function should be computed for a system in
contact with a heat bath at rest, not at the speed of
light~\cite{Elser,SDLCQ}, by use of $e^{-\beta E}$, not 
$e^{-\beta P^-}$.

The standard result for the Casimir force comes from
the computation of the expectation value for the
energy density, which can be written as a sum over
zero-point energies
\be \label{eq:stdresult}
\langle{\cal H}\rangle=\frac{1}{2L}\sum_{n=-\infty}^\infty \int \frac{d^2p_\perp}{(2\pi)^2}E_n,
\ee
with $E_n=\sqrt{p_\perp^2+\left(\frac{2\pi n}{L}\right)^2}$.
The sum can be regulated by a heat-bath factor $e^{-\Lambda E_n}$,
to obtain 
\be
\langle{\cal H}\rangle=\frac{3}{2\pi^2 \Lambda^4}-\frac{\pi^2}{90 L^4}.
\ee
The second term is the regulator-independent effective potential
and determines the Casimir force.  We now consider how
this can be done in light-front coordinates.

%%%%%%%%%%%%%%%%%%%%%%%%%%%%%%%%%%%%%%%%%%%%%%%%%%%%%%%%%%%%%%5
\section{Light-front analysis}  \label{sec:analysis}
%%%%%%%%%%%%%%%%%%%%%%%%%%%%%%%%%%%%%%%%%%%%%%%%%%%%%%%%%%

Our goal is to impose periodic boundary conditions
on a neutral massless scalar field and compute the vacuum energy density.
The light-front mode expansion for the scalar field is
\be \label{eq:mode}
\phi=\int \frac{d\ub{p}}{\sqrt{16\pi^3 p^+}}
   \left\{ a(\ub{p})e^{-ip\cdot x} + a^\dagger(\ub{p})e^{ip\cdot x}\right\},
\ee
with the modes quantized such that 
\be
[a(\ub{p}),a^\dagger(\ub{p}')]=\delta(\ub{p}-\ub{p}').
\ee

For plates perpendicular to the $z$ axis, placed at
$z=0$ and $z=L$, the periodicity imposed is $\phi(z+L)=\phi(z)$.
In light-front coordinates, this periodicity is
\be
\phi(x^++L,x^--L,\vec{x}_\perp)=\phi(x^+,x^-,\vec{x}_\perp),
\ee
which implies $-p^+L/2+p^-L/2=2\pi n$
or $\frac{p_\perp^2}{p^+}-p^+=\frac{4\pi}{L}n$
with $n$ any integer between $-\infty$ and $\infty$. 
The positive solution of this constraint is
\be
p_n^+\equiv \frac{2\pi}{L}n+\sqrt{\left(\frac{2\pi}{L}n\right)^2+p_\perp^2}.
\ee
Then $n=-\infty$ corresponds to $p^+=0$, and $n=\infty$ to $p^+=\infty$.

A discrete mode expansion can be constructed, with use of
discrete annihilation operators
\be
a_n(\vec{p}_\perp)=\sqrt{\left|\frac{dp^+}{dn}\right|}\;a(p_n^+,\vec p_\perp),
\ee
where $\frac{dp^+}{dn}=\frac{2\pi}{L} \frac{p_n^+}{E_n}$.  For these
operators, the commutation relation becomes
\be
[a_n(\vec{p}_\perp),a_{n'}^\dagger((\vec{p}_\perp)^{\,\prime})]
    =\delta_{nn'}\delta(\vec{p}_\perp-\vec{p}_\perp^{\,\prime}).
\ee
The integration over $p^+$ becomes a sum over $n$:
\be
\int dp^+ =\int \frac{dp^+}{dn}dn\rightarrow \sum_n \frac{dp^+}{dn}.
\ee
These yield
\be
\phi(x^+=0)=\frac{1}{\sqrt{2L}}\sum_n \int \frac{d^2p_\perp}{2\pi\sqrt{E_n}}
   \left\{ a_n(\vec{p}_\perp)e^{-ip_n^+x^-/2+i\vec{p}_\perp\cdot\vec{x}_\perp}
  + a_n^\dagger(\vec{p}_\perp)e^{ip_n^+x^-/2-i\vec{p}_\perp\cdot\vec{x}_\perp}\right\}.
\ee
The leading $\frac{1}{\sqrt{2L}}$ factor is consistent with normalization on 
the interval $[-2L,0]$ in $x^-$.

For the free scalar, the light-front energy and longitudinal momentum
densities are ${\cal H}^-=\frac12|\vec\partial_\perp\phi|^2$
and ${\cal H}^+=2|\partial_-\phi|^2$.  Their vacuum expectation
values are
\bea
\langle0|{\cal H}^-|0\rangle&=&\frac{1}{4L}\sum_{n,n'}\int 
      \frac{d^2p_\perp d^2p'_\perp}{(2\pi)^2\sqrt{E_n E_{n'}}}
      \vec{p}_\perp\cdot\vec{p}_\perp^{\,\prime}
      \langle0|a_n(\vec{p}_\perp)a_{n'}^\dagger(\vec{p}_\perp^{\,\prime})|0\rangle \\
      &=&\frac{1}{4L}\sum_n \int \frac{d^2p_\perp}{(2\pi)^2 E_n}p_\perp^2, \nonumber 
\eea
\bea
\langle0|{\cal H}^+|0\rangle&=&\frac{2}{2L}\sum_{n,n'}\int 
      \frac{d^2p_\perp d^2p'_\perp}{(2\pi)^2\sqrt{E_n E_{n'}}} \frac{p_n^+ p_{n'}^+}{4}
      \langle0|a_n(\vec{p}_\perp)a_{n'}^\dagger(\vec{p}_\perp^{\,\prime})|0\rangle \\
      &=&\frac{1}{4L}\sum_n \int \frac{d^2p_\perp}{(2\pi)^2 E_n} (p_n^+)^2. \nonumber 
\eea

The energy density, relative to light-cone coordinates, is
\bea
{\cal E}_{\rm LF}&\equiv&\frac12(\langle0|{\cal H}^-|0\rangle+\langle0|{\cal H}^+|0\rangle) \\
&=& \frac{1}{8L}\sum_n \int \frac{d^2p_\perp}{(2\pi)^2 E_n} (2E_n^2+2\frac{2\pi}{L}nE_n). \nonumber 
\eea
The second term is zero, because it is proportional
to $\sum_{n=-\infty}^\infty n=0$, leaving
\be
{\cal E}_{\rm LF}=\frac{1}{4L}\sum_n\int \frac{d^2p_\perp}{(2\pi)^2}E_n.
\ee
However, we still need to relate this to the energy density ${\cal E}$
relative to equal-time coordinates.

An integration over a finite volume between the plates yields
\be
{\cal E}=\frac{1}{LL_\perp^2}\int_{-2L}^0 dx^- \int_0^{L_\perp} d^2x_\perp {\cal E}_{\rm LF}.
\ee
A change of variable from $x^-$ to $z=(x^+ + x^-)/2$ at fixed $x^+$ brings
\be
{\cal E}=\frac{1}{LL_\perp^2}\int_0^L 2dx^- \int_0^{L_\perp} d^2x_\perp {\cal E}_{\rm LF}=2{\cal E}_{\rm LF}.
\ee
Thus, the energy density is
\be
{\cal E}=\frac{1}{2L}\sum_n\int \frac{d^2p_\perp}{(2\pi)^2}E_n,
\ee
which matches exactly the standard result. It can  
be regulated with the same heat-bath factor $e^{-\Lambda E_n}$, 
as appropriate for a system in contact with a heat bath at rest in an inertial frame.

For the transverse case, where the plate are separated in a direction transverse
to the $z$ axis, the direct implementation of
light-front coordinates by Lenz and Steinbacher does yield the correct result.
This is not surprising, because plates separated
  in the transverse direction can be at rest in an inertial frame.
However, there could be concern that the additional steps 
introduced here will somehow destroy this agreement, and we
need to check that a consistent result is still obtained.

Let the periodicity be in the $x$ direction, with the plates
at $x=0$ and $x=L_\perp$,
to require $\phi(x^+,x^-,x+L_\perp,y)=\phi(x^+,x^-,x,y)$.
The momentum component $p_x$ is then restricted to the discrete
values $p_n\equiv 2\pi n/L_\perp$.  
We define discrete annihilation operators
\be
a_n(p^+,p_y)=\sqrt{\frac{2\pi}{L}}a(p^+,p_n,p_y),
\ee
with the commutation relation
\be
{[}a_n(p^+,p_y),a_{n'}^\dagger(p^{\prime +},p'_y]
=\delta_{nn'} \delta(p^+-p^{\prime +}) \delta(p_y-p'_y).
\ee
The scalar field is again a discrete sum
\bea
\phi(x^+=0)&=&\frac{1}{\sqrt{L_\perp}}\sum_n \int \frac{dp^+ dp_y}{\sqrt{8\pi^2 p^+}}
   \left\{ a_n(p^+,p_y)e^{-ip^+x^-/2+ip_n x+ip_y y} \right. \\
 && \left.+ a_n^\dagger(p^+,p_y)e^{ip^+x^-/2-ip_n x -i p_y y}\right\}. \nonumber
\eea
The leading factor is consistent with the normalization on the interval
$[0,L_\perp]$ in $x$.
The energy density is again constructed from the sum of the minus and
plus components
\bea
\langle0|{\cal H}^-|0\rangle&=&\frac{1}{2L_\perp}\sum_{nn'}
  \int\frac{dp^+dp_ydp^{\prime+}dp'_y}{8\pi^2\sqrt{p^+p^{\prime+}}}
  (p_n p_{n'}+p_y p'_y)   \\
 &&\times
    \langle0|a_n(p^+,p_y)a_{n'}^\dagger(p^{\prime+},p'_y)|0\rangle \nonumber \\
 && =\frac{1}{2L_\perp}\sum_n \int\frac{dp^+ dp_y}{8\pi^2} \frac{p_n^2+p_y^2}{p^+} \nonumber
\eea
and
\bea
\langle0|{\cal H}^+|0\rangle&=& \frac{2}{L_\perp}\sum_{nn'}
   \int\frac{dp^+dp_ydp^{\prime+}dp'_y}{8\pi^2\sqrt{p^+p^{\prime+}}}
   \frac{p^+ p^{\prime +}}{4}\langle0|a_n(p^+,p_y)a_{n'}^\dagger(p^{\prime+},p'_y)|0\rangle \\
 &&  =\frac{1}{2L_\perp}\sum_n \int\frac{dp^+ dp_y}{8\pi^2} p^+.  \nonumber
\eea
Averaged together, these fix ${\cal E}_{\rm LF}$ to be
\be
{\cal E}_{\rm LF}=\frac{1}{2L_\perp}\sum_n \int \frac{dp^- dp^+ dp_y}{8\pi^2}
   \frac{p^- + p^+}{2} \delta\left(p^--\frac{p_n^2+p_y^2}{p^+}\right).
\ee

The delta function is equivalent to the mass-shell condition
\be
\delta(p^--(p_n^2+p_y^2)/p^+)=p^+\delta(p^2)=p^+\delta(E^2-E_n^2), 
\ee
with $E_n=\sqrt{\left(\frac{2\pi}{L_\perp}n\right)^2 +p_z^2+p_y^2}$.
The form of the delta function motivates a conversion to the equal-time variables
$E=(p^+ +p^-)/2$ and $p_z=(p^+-p^-)/2$, which yields 
\be
{\cal E}_{\rm LF}=\frac{1}{2L_\perp} \sum_n \int \frac{2dE dp_z dp_y}{8\pi^2}
   E(E+p_z)\frac{1}{2E_n}\delta(E-E_n).
\ee
The $E$ integral can be done trivially.
The contribution from the $p_z$ term is zero, because that part of the $p_z$
integral is odd.  This yields the
same result as a calculation of the light-front energy density alone; 
the difference is just the contributions proportional to $p_z$,
which integrate to zero.  

This determines the energy density relative to light-front
coordinates as
\be
{\cal E}_{\rm LF}=\frac{1}{4L_\perp}\sum_n\int\frac{dp_z dp_y}{(2\pi)^2}E_n.
\ee
The transformation to the energy density relative to equal-time coordinates
is again just multiplication by two.  Therefore, we find in the
transverse case
\be
{\cal E}=\frac{1}{2L_\perp}\sum_n\int\frac{dp_z dp_y}{(2\pi)^2} E_n,
\ee
which matches the usual equal-time result and is of the same form
as in the longitudinal case.

%%%%%%%%%%%%%%%%%%%%%%%%%%%%%%%%%%%%%%%%%%%%%%%%%%%%%%%%%%
\section{Summary}  \label{sec:summary}
%%%%%%%%%%%%%%%%%%%%%%%%%%%%%%%%%%%%%%%%%%%%%%%%%

We have computed the Casimir effect for parallel plates
in light-front coordinates and obtained the standard result,
by remaining true to the physics of plates at rest.  This is not a
simple constraint in light-front coordinates, but is physically correct. 
Also important for the calculation was the focus on the
equal-time energy as the true vacuum energy that determines
the effective potential and therefore the Casimir force.
This new derivation of the Casimir effect demonstrates
that the physics of the effect is independent of the 
coordinate choice, as it must be, and that light-front 
quantization is not deficient in its treatment of 
such vacuum effects.  The derivation provides additional confidence 
in the usefulness and applicability of the light-front approach.

%%%%%%%%%%%%%%%%%%%%%%%%%%%%%%%%%%%%%%%%%%%%%%%%%%
\acknowledgments
This work was done in collaboration with S.S. Chabysheva
and supported in part by the US Department of Energy.


\begin{thebibliography}{3}

\bibitem{Casimir} Casimir HBG (1948)
On the attraction between two perfectly conducting plates.
Proc.\ K.\ Ned.\ Akad.\ Wet.\ 51: 793-795

\bibitem{LFCasimir} Chabysheva SS, Hiller JR (2013)
Light-front analysis of the Casimir effect.
Phys.\ Rev.\ D 88: 085006

\bibitem{Lenz} Lenz F, Steinbacher D (2003)
The Casimir effect on the light cone.
Phys.\ Rev.\ D 67: 045010

\bibitem{Almeida} Almeida T, Alves VS, Alves DT, Perez S, Rodrigues PLM (2013)
Light front Casimir effect.
Phys.\ Rev.\ D 87: 065028

\bibitem{Dirac} Dirac PAM (1949)
Forms of relativistic dynamics. 
Rev.\ Mod.\ Phys.\ 21: 392-399

\bibitem{DLCQreviews} For reviews of light-cone quantization, see
Burkardt M (2002)
Light front quantization.
Adv.\ Nucl.\ Phys.\ 23: 1-74;
Brodsky SJ, Pauli H-C, Pinsky SS (1998)
Quantum chromodynamics and other field theories on the light cone.
Phys.\ Rep.\ 301: 299-486

\bibitem{Vary} Harindranath A, Vary JP (1988)
Variational calculation of the spectrum of two-dimensional
$\phi^4$ theory in light-front field theory.
Phys.\ Rev.\ D 37: 3010--3013

\bibitem{Elser}  Elser S, Kalloniatis AC (1996)
QED in (1+1)-dimensions at finite temperature: A Study with light cone quantization.
Phys.\ Lett.\ B 375: 285-291

\bibitem{SDLCQ} Hiller JR, Proestos Y, Pinsky S, Salwen N (2004)
$N=(1,1)$ super Yang-Mills theory in 1+1 dimensions at finite temperature.
Phys.\ Rev.\ D 70: 065012

\end{thebibliography}
\end{document}